\documentclass{Interspeech2024}

\interspeechcameraready 

\title{ASGIR: Audio Spectrogram Transformer Guided Classification And Information Retrieval For Birds}

\name[]{Yashwardhan}{Chaudhuri*}
\name[]{Paridhi}{Mundra*}
\name[]{Arnesh}{Batra*}
\name[]{Orchid Chetia}{Phukan}
\name[]{Arun Balaji}{Buduru}

\address{
  IIIT-Delhi, India\\
  *Equal Contributors
}
\email{\{yashwardhan20417, paridhi20392, arnesh23129, orchidp, arunb\}@iiitd.ac.in}

\keywords{speech recognition, human-computer interaction, bird-sound, information retrieval}

\begin{document}

\maketitle

\begin{abstract}
    Recognition and interpretation of bird vocalizations are pivotal in ornithological research and ecological conservation efforts due to their significance in understanding avian behaviour, performing habitat assessment and judging ecological health. This paper presents an audio spectrogram-guided classification framework called ASGIR for improved bird sound recognition and information retrieval.  Our work is accompanied by a simple-to-use, two-step information retrieval system that uses geographical location and bird sounds to localize and retrieve relevant bird information by scraping Wikipedia page information of recognized birds. ASGIR offers a substantial performance on a random subset of 51 classes of Xeno-Canto dataset Bird sounds from European countries with a median of 100\% performance on F1, Precision and Sensitivity metrics. Our code is available as follows: https://github.com/MainSample1234/AS-GIR .
\end{abstract}

\section{Introduction}
Bird vocalizations play a vital role in ornithological research and ecological conservation efforts, aiding in understanding avian behaviour, assessing habitats, and evaluating ecological health\cite{sprengel2016audio}. However, accurately recognizing and interpreting bird sounds pose significant challenges due to diverse vocalizations across species and environments.
Traditional methods of bird sound analysis often rely on manual observation and classification by ornithologists or usage can be time-consuming and inefficient\cite{michaud2023unsupervised}.To address these limitations, recent advancements in machine learning and signal processing have shown promise in automating bird sound recognition \cite{swaminathan2024multi}\cite{9747202}.\\
This paper introduces \textbf{ASGIR} (\textbf{A}udio \textbf{S}pectrogram Transformer \textbf{G}uided
Classification and \textbf{I}nformation \textbf{R}etrieval for Birds), a framework accompanied by a UI platform for bird sound recognition and information retrieval. ASGIR shows improved Precision, Recall and F1 scores in bird sound classification across diverse avian species.
Our platform is a two-step information retrieval system that
integrates geographical location data with bird sound classification. This system enables the localization and retrieval of relevant bird information by scraping Wikipedia pages for recognized bird species.\\
The experimental results of ASGIR show a substantial performance on a random subset of 51 classes of Xeno-Canto dataset Bird sounds from European countries with an impressive median performance of 100\% Precision, Recall and F1.

\section{Application Overview}
ASGIR leverages bird sound audio and specified recording areas to predict and extract bird-related information from Wikipedia pages. Initially, users provide a recording area, effectively narrowing the search scope. Subsequently, ASGIR employs its audio classifier to identify bird sounds within the designated region. The system then scrapes relevant data from Wikipedia pages corresponding to the identified bird species.
\textbf{Model:} As described in Figure 1. Our approach uses an Audio Spectrogram Transformer\cite{gong21b_interspeech}, pre-trained on AudioSet dataset\cite{7952261} as an upstream feature extractor for generating embeddings of given bird sound. We later use a multi-class Support Vector Machine(SVM)\cite{708428} as a downstream classifier to classify the bird from the 51 possible categories.\\
\textbf{Training:} Our dataset from the Xeno-Canto database contains 153 recordings from 51 classes of birds specified in Table 1, with an average audio length of 45 seconds. Each recording is split into 2-second samples to generate 4855 recordings for training and testing. We use 80\% - 20\% split for training and testing. We use a standard linear kernel to train SVM and AST\cite{gong21b_interspeech} for feature extraction\\
\textbf{Web Scraping:} Once the bird has been identified through its audio signature and the location data supplied by the user, ASGIR proceeds to extract relevant information from Wikipedia pages. This process involves employing web scraping techniques to parse the HTML tags on the webpage, thereby retrieving comprehensive details about the bird. This information typically encompasses the bird's typical habitats, distinctive characteristics, and other relevant attributes.

\begin{figure*}[h]
    \centering
    \includegraphics[width=\textwidth]{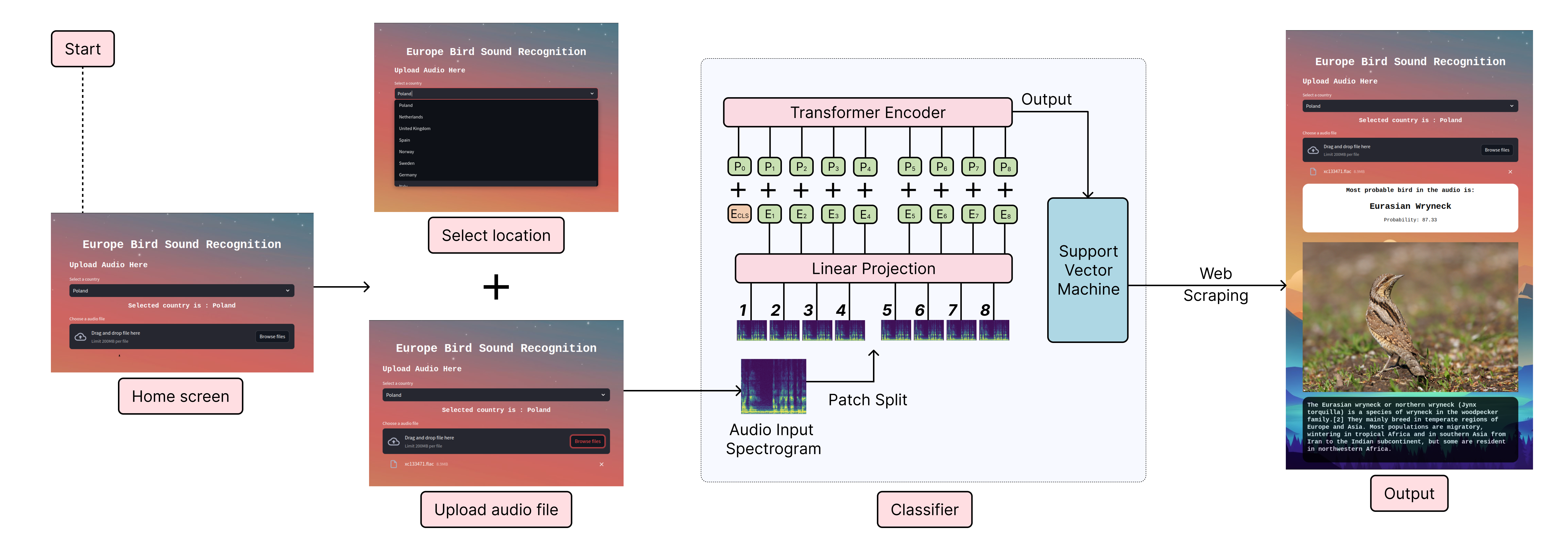}
    \caption{\textbf{ASGIR Workflow:} The user starts by entering 1. audio recording and  2.) location of the audio recording from a drop-down menu. The model narrows the search space based on the location of the recording and then runs the ASGIR audio classifier to detect bird names. We use the bird name to scrape information about its habitat and characteristics by parsing HTML tag information from Wikipedia pages.}
    \label{fig:example}
\end{figure*}

\section{Evaluation}
\textbf{Class-wise performance: } We evaluate the class-wise performance of ASGIR on F-1, Precision and Recall Metrics, as shown in Table 2. We find that our Information Retrieval system has an impressive median performance of 100\% Precision, Recall and F1 scores.\\
\begin{table}[!ht]
    \centering
    \renewcommand{\arraystretch}{1.1}
    \resizebox{0.5\textwidth}{!}{
    \begin{tabular}{|l|l|l|l|l|}
        \hline
        \textbf{Class} & \textbf{Precision} & \textbf{Recall} & \textbf{F1-Score} & \textbf{Support} \\ \hline
        Barn-Swallow & 1.00 & 1.00 & 1.00 & 22 \\ 
        Black-headed-Gull & 1.00 & 1.00 & 1.00 & 25 \\ 
        Common-Chiffchaff & 1.00 & 0.92 & 0.96 & 26 \\ 
        Common-Linnet & 0.97 & 0.94 & 0.95 & 31 \\ 
        Common-Moorhen & 1.00 & 0.96 & 0.98 & 28 \\
        Common-Nightingale & 0.95 & 0.90 & 0.93 & 21 \\ 
        Common-Swift & 1.00 & 1.00 & 1.00 & 14 \\ 
        Common-Whitethroat & 0.86 & 1.00 & 0.92 & 12 \\ 
        Common-Wood-Pigeon & 1.00 & 1.00 & 1.00 & 10 \\ 
        Corn-Bunting & 1.00 & 1.00 & 1.00 & 51 \\
        Eurasian-Blue-Tit & 0.94 & 1.00 & 0.97 & 15 \\
        Eurasian-Jay & 0.92 & 0.92 & 0.92 & 12 \\
        Eurasian-Magpie & 1.00 & 1.00 & 1.00 & 17 \\ 
        Eurasian-Oystercatcher & 1.00 & 1.00 & 1.00 & 20 \\
        Eurasian-Reed-Warbler & 1.00 & 1.00 & 1.00 & 28 \\
        Eurasian-Skylark & 0.99 & 0.99 & 0.99 & 67 \\
        Eurasian-Wren & 1.00 & 0.93 & 0.97 & 15 \\ 
        Eurasian-Wryneck & 1.00 & 1.00 & 1.00 & 38 \\ 
        European-Golden-Plover & 1.00 & 1.00 & 1.00 & 12 \\ 
        European-Nightjar & 0.96 & 1.00 & 0.98 & 22 \\ 
        European-Turtle-Dove & 1.00 & 0.85 & 0.92 & 13 \\ 
        Garden-Warbler & 1.00 & 0.96 & 0.98 & 25 \\ 
        Grey-Partridge & 0.92 & 0.86 & 0.89 & 14 \\ 
        Meadow-Pipit & 0.93 & 1.00 & 0.97 & 28 \\ 
        Northern-Lapwing & 1.00 & 1.00 & 1.00 & 19 \\ 
        Northern-Raven & 0.80 & 0.97 & 0.88 & 29 \\ 
        Red-throated-Loon & 1.00 & 1.00 & 1.00 & 10 \\ 
        River-Warbler & 1.00 & 1.00 & 1.00 & 23 \\ 
        Sedge-Warbler & 1.00 & 0.98 & 0.99 & 63 \\ 
        Stock-Dove & 1.00 & 1.00 & 1.00 & 22 \\ 
        Western-Jackdaw & 1.00 & 1.00 & 1.00 & 18 \\ 
        Willow-Ptarmigan & 1.00 & 1.00 & 1.00 & 28 \\ 
        Willow-Warbler & 1.00 & 0.90 & 0.95 & 21 \\
        \hline
        
    \end{tabular}
    }
    \caption{Macro F1, Precision and Recall performance of Popular Eurasian birds in Xeno-Carto Database subset}
\end{table}
\textbf{Ablations} As Shown in Table 1. We performed an ablation study with the trial of different upstream and downstream networks, where we found that the AST + SVM configuration was performing better than its counterparts. We also found that ASGIR performs marginally better when provided with a location-based search, giving an increase of approximately 0.5\% in accuracy. 
\begin{table}[!ht]
    \centering
    \resizebox{0.45\textwidth}{!}{
    \begin{tabular}{lllll}
    \hline
        \textbf{Model} & \textbf{F1-Score Macro} & \textbf{Precision} & \textbf{Recall} & \textbf{Accuracy} \\ \hline
        AST-1DCNN & 0.965 & 0.967 & 0.965 & 0.973 \\ \hline
        AST-SVM & 0.972 & 0.974 & 0.972 & 0.975 \\ \hline
        AST-GMM & 0.933 & 0.917 & 0.965 & 0.967 \\ \hline
        \textbf{ASGIR- Ours} & \textbf{0.973} & \textbf{0.979} & \textbf{0.973} & \textbf{0.984} \\ \hline
    \end{tabular}
    }
    \caption{Comparison of model performance metrics. GMM: Gaussian Mixture Model, SVM: Support Vector Machine, 1D-CNN: 1-Dimensional CNN}
    \label{tab:model-performance}
\end{table}
\section{Conclusion}

In conclusion, ASGIR presents a robust framework for bird sound recognition and information retrieval, addressing the significant challenges in ornithological research and ecological conservation efforts. ASGIR improves Precision, Recall, and F1 scores in bird sound classification by leveraging audio spectrogram-guided classification and web scraping techniques. Additionally, its two-step information retrieval system effectively localizes and retrieves relevant bird information from Wikipedia pages based on geographical location and bird sounds. The experimental results demonstrate ASGIR's performance across avian species, showcasing its potential utility in various ecological and ornithological applications.
\bibliographystyle{IEEEtran}
\bibliography{main}

\end{document}